\documentclass[twocolumn]{aastex63}

\DeclareRobustCommand{\Figref}[1]{Fig.~\ref{#1}}

\usepackage{flushend}
\usepackage{amsmath}
\graphicspath{{./figures/}}

\begin{document}

% Title
\title{Progenitor with small reaction networks should not be used as
  initial conditions for core collapse}

\author[0000-0002-6718-9472]{M.~Renzo}
\affiliation{University of Arizona, Department of Astronomy \& Steward Observatory, 933 N.~Cherry Ave., Tucson, AZ 85721, USA}

\author[0000-0003-1012-3031]{J.~A.~Goldberg}
\affiliation{Center for Computational Astrophysics, Flatiron Institute, 162 5th Avenue, New York, NY 10010, USA}

\author[0000-0002-2215-1841]{A.~Grichener}
\affiliation{Department of Physics, Technion, Kiryat Hatechnion, Haifa 3200003, Israel}

\author[0000-0003-3115-2456]{O.~Gottlieb}
\affiliation{Center for Computational Astrophysics, Flatiron Institute, 162 5th Avenue, New York, NY 10010, USA}

\author[0000-0002-8171-8596]{M.~Cantiello}
\affiliation{Center for Computational Astrophysics, Flatiron Institute, 162 5th Avenue, New York, NY 10010, USA}

\begin{abstract}
  \noindent
  Core collapse initial conditions are a bottleneck in understanding
  the explosion mechanism(s) of massive stars. Stellar evolution codes
  struggle after carbon burning, and either stop or adopt numerical
  simplifications missing crucial physics. The use of small nuclear
  reaction networks (NRN) that account for energy production but
  bypass weak reactions is typical, but insufficient to study the
  dynamics of the collapse. We advise against the use of progenitors
  computed with small NRN in expensive multidimensional simulations of
  core collapse, bounce, (jet formation), and explosion.
\end{abstract}

\section{Modeling uncertainties compound}

Massive stars ($\gtrsim 7-10\,M_\odot$, \citealt{doherty:15,
  poelarends:17}) end their life collapsing and possibly exploding
\citep[e.g.,][]{janka:12, burrows:21, soker:24}. Although a consensus
around the ``neutrino-driven'' paradigm is establishing
\citep[e.g.,][]{wang:23, nakamura:24}, details remain debated
\citep[e.g.,][]{shishkin:22, soker:22}, especially for magnetic
fields and rotation \citep[e.g.,][]{symbalisty:84, mosta:15, aloy:21}.

The collapse/explosion is sensitive to the initial conditions
\citep[e.g.,][]{ott:18, kuroda:18, burrows:23, nakamura:24},
determined by the late core evolution. Only limited sets of
nonrotating progenitors computed sufficiently late exhist
\citep{woosley:02, sukhbold:16, farmer:16, renzo:17, wang:24} and even
fewer include rotation \citep{heger:00,
  aguilera-dena:18}.

\cite{farmer:16} showed the impact of algorithmic choices on the final
core of nonrotating stars, demonstrating that small
($\sim{}20$-isotope) NRN are insufficient to produce reliable initial
conditions for multidimensional core collapse studies. These NRNs
allow for deleptonization through one single compound reaction (e.g,
$^{56}\mathrm{Fe}+2e^{-}$ $\rightarrow$ $^{56}\mathrm{Cr}+2\nu_{e}$).
This predetermines $Y_e$ \emph{throughout} the core, thus
the effective \cite{chandrasekhar:31} mass $\propto {Y_e}^2$, and
ultimately the outcome of the collapse. This affects stellar evolution
models computed with the \texttt{approx} family of nuclear reaction
networks \citep{timmes:00}. Here, we extend this result to fast
rotating progenitors.

\begin{figure*}[htbp]
  \centering
  \includegraphics[width=0.9\textwidth]{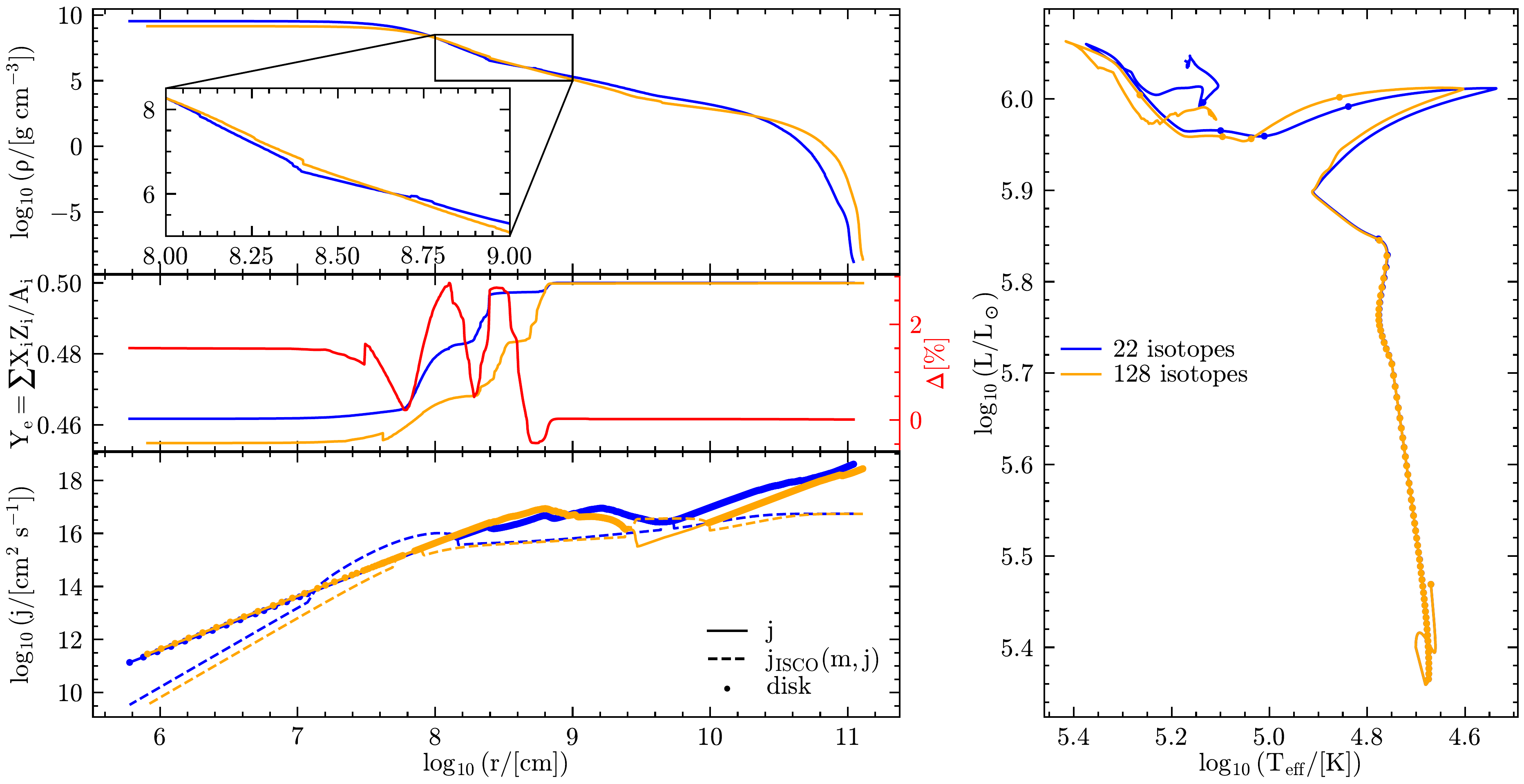}
  \caption{Comparison of chemically homogeneous models at the onset of
    core-collapse. Right: HRD (one dot every $10^5$ years).
    Top left: density. Middle left: $Y_e$, relative difference in red
    (right vertical axis). Bottom left: specific angular momentum $j$.
    The dashed line shows $j_\mathrm{ISCO}$ assuming the
    accretion of the enclosed $m$ and $j$. Dots correspond to regions
    that collapse forming a disk, $j\geq j_\mathrm{ISCO}$.}
  \label{fig:comparison}
\end{figure*}

We ran two $40\,M_\odot$, Z=0.001 stars initially rotating at 60\% of
breakup with MESA \citep[][r24.03.1]{jermyn:23}. A reproducibility
package is available at
\href{https://doi.org/10.5281/zenodo.11375523}{doi:10.5281/zenodo.11375523}.
The models differ only in the NRN, \texttt{mesa\_128} (orange in
\Figref{fig:comparison}), sufficient to follow the deleptonization
\citep{farmer:16}, and \texttt{approx21\_plus\_cr56} (blue). We
include rotational mixing \citep{heger:00}, magnetic torques
\citep{spruit:02}, and hydrodynamics. We expunge spurious velocities
in layers not in sonic contact with the innermost core (CO, Si, or
Fe). We performed resolution tests (up to mesh sizes
$\gtrsim 10\,000$), and found that numerical resolution affects
convection in the inner $5\,M_\odot$ and consequently the inner
structure \citep[e.g.,][]{sukhbold:14, schneider:23}. This emphasizes
the importance of resolution tests \citep{farmer:16, farag:22}. The models we
present have identical resolution requirements and are representative
of the range of structures that can be obtained.

Both models experience blueward rotationally-induced chemically
homogeneous evolution \citep[e.g.,][]{maeder:00} and are computed
until the infall velocity exceeds $300\,\mathrm{km\
  s^{-1}}$.
\Figref{fig:comparison} (right) shows the Hertzsprung-Russell diagram
(HRD). The two tracks are observationally indistinguishable, with most
differences confined to fast evolutionary phases: small NRNs are
sufficient to simulate surface properties.

Nevertheless, the interior structure
significantly different:
the compactness parameter \citep{oconnor:11} of the small (large) NRN
model is $\xi_{2.5}=0.376$ ($0.262$). The $\sim{}44\%$ variation in
$\xi_{2.5}$ is comparable to uncertainties introduced by numerical
resolution \citep{farmer:16}, mass loss \citep{renzo:17}, and
overshooting \citep{davis:19}. No single parameter is
sufficient to characterize ``explodability''
\citep[e.g.,][]{ertl:16, vartanyan:21}, and we show on the left
of \Figref{fig:comparison} the internal profiles.

The top panel shows the density profiles, with an inset focusing on
the region where the success or failure of an explosion is decided
\citep[e.g.,][]{ertl:16, boccioli:23, burrows:23}.

The middle panel shows the $Y_e$ profile. All the mass collapsing
interior to $10^9\,\mathrm{cm}$ shows \emph{structured} relative
variations
$\Delta=(Y_e^\mathrm{small\ NRN}-Y_e^\mathrm{large\
  NRN})/Y_e^\mathrm{large\ NRN}\simeq 1-3\%$ (red, righ vertical
axis). The range of $Y_e$ found at the onset of collapse is $0.4-0.5$,
making this nonnegligible.

The bottom left panel shows the specific angular momentum ($j$, solid
lines) and the innermost stable orbit angular momentum assuming
accretion of the enclosed mass and $j$ ($j_\mathrm{ISCO}$, dashed
lines, \citealt{bardeen:72}). Wherever $j\geq j_\mathrm{ISCO}$ an
accretion disk has to form (thicker dots). In both cases the inner
$\sim{}10^{7}\,\mathrm{cm}$ retain too much angular momentum to
directly collapse, and we expect a proto-neutron-star phase. After the
collapse of this material, the 128-isotope model shows the formation
of a disk immediately, while the 22-isotope model only after
$\gtrsim 1\,\mathrm{s}$. From disk formation onwards the evolution
will diverge because of feedback processes (disk wind, jet,
\citealt{gottlieb:22}). This difference occurs because of the
diverging evolution of the density and thus $j$ caused by the
treatment of nuclear physics.

Core collapse is a $\sim{}1\%$ problem: only this fraction of the
gravitational potential released
needs to be ``harvested''
to produce a successful supernova. Predictive
simulations require high accuracy initial conditions. Many uncertain
processes occur in massive stars \citep[e.g.,][]{woosley:02,
  langer:12}, and modeling choices matter, including resolution
\citep{farmer:16}, mass loss \citep{renzo:17}, mixing
\citep{davis:19}, spurious envelope velocities
\citep[e.g.,][]{farmer:16, aguilera-dena:18}, and dimensionality of
the simulations \citep[e.g.,][]{fields:22}. We advise against running
expensive multidimensional simulations of core-collapse with initial
conditions that are known to not include enough physics to address the
questions motivating such simulations in the first place. This means
not using models computed only until carbon depletion, models wiht
small NRN, and/or underresolved
models.

At fixed angular momentum transport, small NRN result in significantly
different rotationally-powered explosions for the same core-collapse
physics, making robust conclusions impossible.

Progenitors computed with small NRN remain useful for studying the
outer layers of stars up to collapse, and possibly to make relative
statements, but require care to marginalize any conclusion against
this large systematic modeling uncertainty in the progenitor.

\bibliography{./CHE_networks.bib}
\end{document}